\documentclass[preprint2]{aastex}

\shorttitle{Integrated spectra with BaSTI}
\shortauthors{Percival et al.}

\begin{document}


\title{A large stellar evolution database for population synthesis studies. IV. 
Integrated properties and spectra. }


\author{Susan M. Percival and Maurizio Salaris}
\affil{Astrophysics Research Institute, Liverpool John Moores University, Twelve 
Quays House, Birkenhead, CH41 1LD, UK}
\email{smp@astro.livjm.ac.uk,ms@astro.livjm.ac.uk}

\and

\author{Santi Cassisi and Adriano Pietrinferni}
\affil{INAF-Osservatorio Astronomico di Collurania, Via M. Maggini, I-64100 Teramo, Italy}
\email{pietrinferni@oa-teramo.inaf.it,cassisi@oa-teramo.inaf.it}






\begin{abstract}
This paper is the 4th in a series describing the latest additions to the 
BaSTI stellar evolution database, which consists of a large set of 
homogeneous models and tools for population synthesis studies, covering
ages between 30~Myr and $\sim$20~Gyr and 11 values of $Z$ (total metallicity).
Here we present a new set of low and high resolution synthetic spectra based 
on the BaSTI stellar models, covering a large range of simple stellar 
populations (SSPs) for both scaled solar and $\alpha$-enhanced metal 
mixtures.  This enables a completely consistent study of 
the photometric and spectroscopic properties of both resolved and
unresolved stellar populations, and allows us to make detailed tests on
specific factors which can affect their integrated properties.
Our low resolution spectra are suitable for deriving broadband 
magnitudes and colors in any photometric system.  These spectra cover the
full wavelength range (9--160000nm) and include all evolutionary stages up 
to the end of AGB evolution.  
Our high resolution spectra are suitable for studying the 
behaviour of line indices and we have tested them against a large sample of 
Galactic globular clusters.  We find that the range of ages, iron 
abundances [Fe/H], and degree of $\alpha$-enhancement predicted by the models
matches the observed values very well. 
We have also tested the global consistency of the BaSTI models by making 
detailed comparisons between ages and metallicities derived from isochrone 
fitting to observed colour-magnitude diagrams, and from line index strengths, 
for the Galactic globular cluster 47~Tuc and the open cluster M67.  
For 47~Tuc we find reasonable agreement between the 2 methods, within the
estimated errors.  From the comparison with M67 we find non-negligible 
effects on derived line indices caused by statistical 
fluctuations, which are a result of the specific method used to populate an 
isochrone and assign appropriate spectra to individual stars. 
\end{abstract}





\section{Introduction}
\label{sec:intro}

Large grids of stellar models and isochrones are necessary tools to interpret
photometric and spectroscopic observations of resolved and unresolved 
stellar populations. This, in turn, allows one to both test the theory of 
stellar evolution, and investigate the formation and evolution of galaxies,
one of the major open problems of modern astrophysics.

To this purpose, we started in 2004 a major project aimed at creating a
large and homogeneous database of stellar evolution models and isochrones
(BaSTI -- a Bag of Stellar Tracks and Isochrones) that cover
a large chemical composition range relevant to stellar populations
in galaxies of various morphological types. In addition, BaSTI includes
the option of choosing between different treatments of core convection and
stellar mass loss employed in the model calculations.
In its present form, BaSTI contains a grid of models that cover
stellar populations with ages ranging between 30~Myr and $\sim$20~Gyr, 
which include all evolutionary
phases from the Main Sequence (MS) to the end of the Asymptotic
Giant Branch (AGB) evolution or carbon ignition, depending on the
value of stellar mass.  They are calculated for two metal
mixtures (scaled solar and $\alpha$-enhanced) and 11 values of $Z$ 
(total metallicity), each with two choices for the Reimers mass loss 
parameter $\eta$ \citep{reimers}, and two mixing prescriptions during the
Main Sequence (without and with overshooting from the Schwarzschild 
boundary). 
Broadband magnitudes and colors for the BaSTI models and isochrones
are provided in various photometric systems (Johnson-Cousins, Sloan, 
Str\"omgren, Walraven, ACS/$HST$), making use of bolometric corrections 
derived from an updated set of model atmosphere calculations, for 
element mixtures consistent with those employed in the stellar evolution 
computations\footnote{All the results presented in this work are
based on the 2008 release of the BaSTI archive. We refer to the official
BaSTI website for more details on this new release.}.

In addition to stellar models and isochrones, BaSTI also provides a series 
of Web tools that enable an interactive access to the database and make it 
possible to compute
user-specified evolutionary tracks, isochrones, stellar luminosity functions
(star counts in a stellar population as a function of magnitude) plus
synthetic color-magnitude diagrams (CMDs) for arbitrary star formation
histories (SFHs).
The results of this major effort have been published in 
\citet[][ hereinafter Papers~I and II]{basti1,basti2} and 
\citet[][hereinafter Paper~III]{basti3}.
All Web tools, models and isochrones can be found at the BaSTI official
website
\url{http://193.204.1.62/index.html}.

The database has been extensively tested against
observations of local stellar populations and eclipsing binary systems
(see, for some examples, the tests presented in Papers~I,
~II and ~III; \citealt{ms07,tomasella}).
In its present form BaSTI can be used to investigate resolved stellar 
populations, and indeed it has been widely employed in studies of Galactic 
and extragalactic resolved star clusters 
\citep[see, e.g.][for just a few examples]{deangeli,villanova,mackey} and 
in the determination of SFH and chemical enrichment history of resolved
Local Group galaxies \citep[see, e.g.][]{gallart,barker,gulli,carrera}.

Integrated colors and magnitudes for simple (single-age, single metallicity) 
stellar populations (SSPs) can be easily determined from the isochrones. 
The only additional pieces of information needed are an Initial Mass Function 
(IMF) and the relationship between the initial and final (remnant) stellar 
masses.  Indeed, some integrated colors from BaSTI models have 
already been tested against empirical constraints in Papers~II and III.
Moreover, \citet{paj} and \citet{salcas} have already
employed integrated colors from BaSTI isochrones to address issues 
related to the SFH of unresolved dwarf and elliptical galaxies, 
and extragalactic globular cluster ages.

A further development of BaSTI, which we present in this paper, is the 
inclusion of integrated spectra (including spectral line indices) 
for SSPs spanning the same parameter space -- in terms of ages, metallicities
and heavy element mixtures -- covered by the isochrones of Papers~I and II.
Corresponding quantities for composite stellar populations with an arbitrary
SFH can then be easily determined by integration over a range of discrete SSPs.

We present both low and high resolution 
versions suited to different purposes.  Our low resolution spectra cover 
all evolutionary stages and are based on libraries of (predominantly) 
theoretical stellar spectra, consistent with those used to produce the 
bolometric corrections employed in the isochrone calculations.  
This ensures that our SSP integrated broadband magnitudes and colors 
obtained from either adding up the broadband fluxes of the 
individual stars populating the appropriate isochrone, or from their 
composite integrated spectrum, will be exactly the same (see test in 
section~\ref{sec:lrs}).  
Our high resolution spectra are suitable for deriving line strengths and 
making detailed tests of their behaviour as a function of age and chemical
composition, and also horizontal branch morphology and sampling in the case 
of resolved 
stellar clusters (see tests in sections~\ref{sec:47tuc} and \ref{sec:m67}).

With the inclusion of predictions of both photometric and spectroscopic 
integrated properties of SSPs, BaSTI will provide a well tested,
fully homogeneous and up-to-date set of theoretical tools that enable us to
investigate self-consistently both resolved and unresolved stellar 
populations. We present a global test of this self-consistency and 
reliability of the database, by considering both the observed CMD and 
integrated spectrum of the well-studied clusters 47~Tuc and M67, to 
intercompare ages and metallicities obtained from
our high-resolution spectra with the results from fitting theoretical
isochrones to the observed CMD, and with independent chemical composition 
estimates from spectroscopy of individual stars.

The structure of the paper is as follows.  
A description of the method used to create our theoretical integrated 
spectra is given in Section~\ref{sec:intspec}, including separate 
detailed descriptions of our low and high resolution spectra in 
Sections~\ref{sec:lrs} and \ref{sec:hrs} respectively.
This is followed in Section~\ref{sec:comparison} by a comparison of our 
derived line indices with the Galactic globular cluster sample of 
\citet{schi05}.  A global consistency test of our isochrones, 
integrated colors and spectra using photometric and spectroscopic data for 
47~Tuc and M67 is presented in 
Sections~\ref{sec:47tuc} and \ref{sec:m67}. 
A brief summary in Section~\ref{sec:sum} closes the paper.

\section{Creating integrated spectra:  Method}
\label{sec:intspec}

Before an integrated spectrum can be created, the SSP (or composite 
population) itself must be created by `populating' the relevant isochrone 
(or combination of isochrones).  BaSTI isochrones are defined in terms of 
total metallicity $Z$ (with corresponding [Fe/H] depending on whether 
scaled solar or $\alpha$-enhanced models are being used) and age.  Each 
isochrone consists of 2250 discrete evolutionary points, each of which 
is defined in terms of effective temperature $T_{eff}$, luminosity $L$, 
and mass M (from which log$g$ can also be derived).  An isochrone is 
populated by applying the IMF by \citet{kroupa} with an appropriately 
chosen normalization constant -- the value of this constant is given by
${\rm M_t}=\int_{{\rm M_l}}^{{\rm M_u}} \psi({\rm M}) {\rm M} d{\rm M}$ 
where ${\rm M_t}$ is the total mass of stars formed in the initial burst 
of star formation that originated the SSP. The upper and lower integration 
limits correspond to the lowest and highest mass stars formed during the 
burst, and are fixed to 0.1 and 100$M_{\odot}$, respectively.  We choose 
to set ${\rm M_t}$ equal to 1$M_{\odot}$.
We note here that the lower mass limit of the BaSTI database is 
0.5~$M_{\odot}$, however objects with masses below this limit contribute 
negligibly to the integrated magnitudes and colors, as verified in 
\citet{salcas} by implementing the very low mass star models by 
\citet{cass01}, that extend down to $\sim$0.1$M_{\odot}$.  Hence, stars of 
$<0.5~M_{\odot}$ are included in the total mass and are accounted for in the 
normalization constant, but their photometric properties are not included
in the integrated spectrum.

Once an SSP has been created, each evolutionary point in a particular 
simulation is assigned a spectrum by interpolating linearly in [Fe/H], 
$T_{eff}$ and log$g$ amongst spectra in the appropriate spectral library 
(see Sections~\ref{sec:lrs} and \ref{sec:hrs} for choices of spectral 
libraries).  Each spectrum is scaled by the stellar surface area, via the 
stellar radius derived from the evolutionary point on the isochrone, and 
then weighted appropriately, according to the number of stars predicted by 
the normalized IMF, as described above.  The 2250 individual spectra are 
then summed to obtain the final integrated spectrum.

\subsection{Integrated spectra: low resolution}
\label {sec:lrs}

Our low resolution integrated spectra incorporate all evolutionary stages 
covered by the BaSTI isochrones, as detailed in Section~\ref{sec:intro}.  
This yields a full spectral energy distribution (SED) suitable for 
calculating broad-band colors and, for example, deriving K-corrections.

The majority of the low resolution spectra used in this study are from 
the \citet{castkur} data set\footnote{available at 
http://kurucz.harvard.edu/grids.html}, based on the ATLAS9 model atmospheres.  
These spectra cover effective temperatures, $T_{eff}$, from 3500K up to 
50000K and log$g$ from 0.0 dex to 5.0 dex.  Metallicities range from 
[Fe/H]$=-2.5$ to $+0.5$ for both scaled solar and $\alpha$-enhanced 
mixtures -- the level of $\alpha$-enhancement being fixed at [$\alpha$/Fe]=0.4,
the same as that used in the BaSTI stellar models.  These spectra cover the 
full wavelength range (9--160000nm) and sampling varies from 1nm at 300nm
to 10nm at 3000nm.

For stars with $T_{eff} <$ 3500K (except for carbon stars in the AGB 
phase -- see below) we supplement the Castelli/Kurucz spectra with those 
from the BaSeL 3.1 (WLBC 99) spectral library \citep{wlbc99}.  
The WLBC99 spectra have the same wavelength coverage and sampling as 
the Castelli/Kurucz ones and are provided for scaled solar abundances only, 
in the range $-2.0 \leq {\rm [Fe/H]} \leq +0.5$.  
Since we are only using the WLBC99 spectra for the coolest stars (mostly RGB, 
and some AGB), the lack of $\alpha$-enhanced spectra should not be a problem 
as these stars only contribute significant flux in the reddest part of the 
spectrum where it is known that the colors are generally insensitive to 
the metal distribution \citep[see e.g.][]{alonso,cas04}.

For the AGB carbon stars we use the averaged AGB spectra of \citet{lanmou}.  
These (empirical) spectra cover 510--2490nm, with linear extrapolation to 
zero flux at 350 and 5000nm, at a resolution of 0.5nm (5{\AA}), which we 
rebin to match the Castelli/Kurucz sampling.  They are defined in terms of 
$T_{eff}$ only, as no metallicity information is available.  

\begin{figure}
\plotone{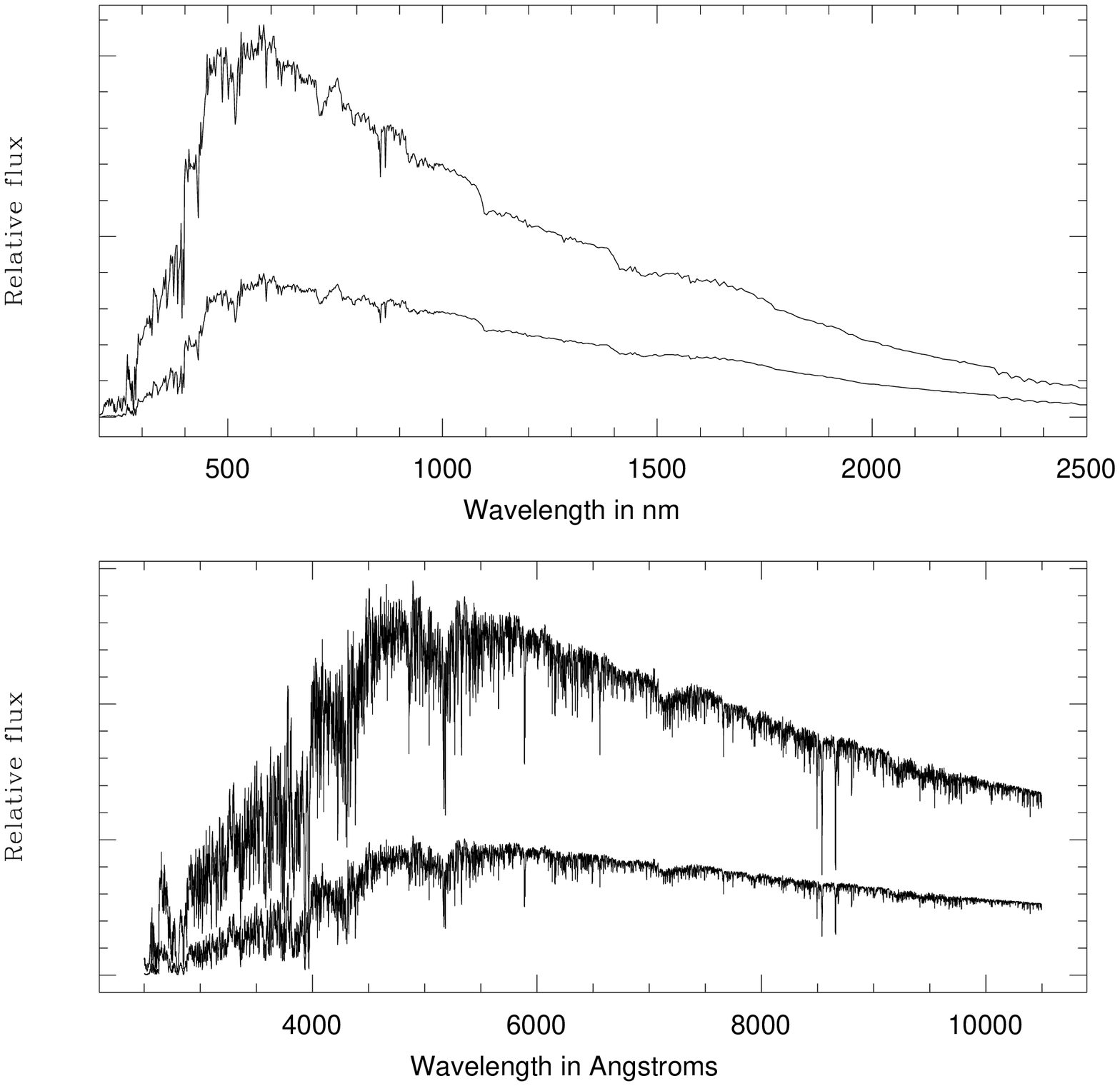}
\caption{Sample low resolution (upper panel) and high resolution (lower panel)
integrated spectra from scaled solar, $Z$=0.0198 (the solar value) models.
Ages displayed are 3 and 10 Gyr, which are the upper and lower spectra 
respectively on each plot.}
\label{sample_spec}
\end{figure}

Some representative examples of our low resolution integrated spectra are 
shown in Figure~\ref{sample_spec}.
Since the low resolution spectra have full wavelength coverage and cover all 
evolutionary points on the isochrones, they are useful for identifying the 
flux contribution from specific stars or evolutionary stages.  This is 
important when we assess the validity of line strengths derived from the high
resolution spectra, for which there is a low temperature cut-off in the 
spectral library used (see section~\ref{sec:hrs}).

\subsubsection{Integrated colors -- behaviour and comparisons}
\label{sec:cols}

As a sanity check, we verified the consistency of the colors derived from
our low resolution integrated spectra with those predicted from the isochrones 
by the BaSTI population synthesis Web tool.  Colors were derived from the 
spectra by convolving the integrated spectra with broadband filter profiles 
and normalising to a synthetic Vega spectrum \citep{vega}.  Colors are 
obtained from an isochrone by adding up the flux contributions from each 
evolutionary point in the relevant broadband filter.  For a given age $t$, 
metallicity $Z$, plus an IMF $\psi({\rm M})$, we simply integrate the 
flux in the generic broadband filter $\lambda$ along the representative 
isochrone, i.e.
$$
M_{\lambda}=-2.5 \ {\rm log} \left(\int_{{\rm M_l}}^{{\rm M_u}}
\psi({\rm M}) 10^{-0.4 M_{\lambda}({\rm M})} d{\rm M}\right)
$$
We performed this test for a typical globular cluster ($Z$=0.001, $t$=10Gyr) 
and a typical open cluster ($Z$=0.0198, $t$=3.5Gyr), from $(U-B)$ through 
to $(J-K)$.  For the solar metallicity cluster, all colors are consistent 
within $\lesssim$ 0.015 mag whilst for the lower metallicity simulation 
the agreement is even better (within $<$ 0.01mag).

BaSTI isochrones currently provide magnitudes for `standard' $UBVRIJHKL$ 
filters.  Since we have shown that the resulting colors are consistent 
with those derived from our integrated spectra, magnitudes and colors in 
other photometric systems can reliably be derived by convolving the spectra 
with the required filter response curves, and using the appropriate 
normalizing spectrum (as described above).

\begin{figure}
\plotone{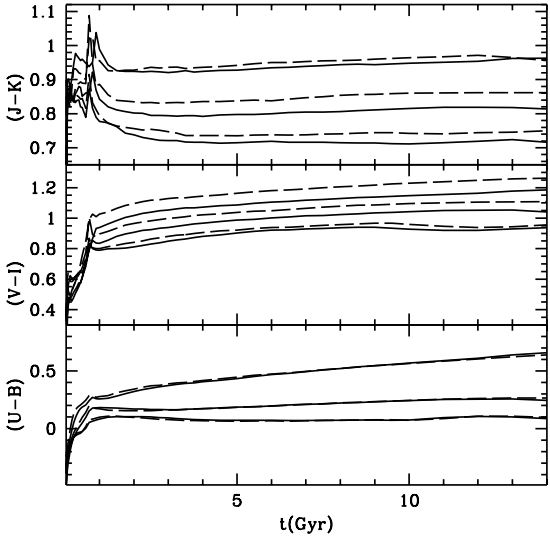}
\caption{SSP colors from BaSTI models.  Solid lines are the scaled solar 
models for [Fe/H]=$+0.06$, $-0.66$ and $-1.27$ (upper, middle and lower 
lines respectively on each plot).  Dashed lines are the 
$\alpha$-enhanced models for [Fe/H]=$+0.05$, $-0.70$ and $-1.31$.}
\label{basticols}
\end{figure}

As an example of our integrated SSP color predictions, we display in 
Figure~\ref{basticols} the run of $(U-B)$, $(B-V)$ and $(J-K)$ as a 
function of the SSP age, for the scaled solar and $\alpha$-enhanced mixtures.  
Scaled solar models are shown for [Fe/H]=$+0.06$, $-0.66$ and $-1.27$, and 
are matched to the $\alpha$-enhanced models with almost exactly the same 
[Fe/H] values ([Fe/H]=$+0.05$, $-0.70$ and $-1.31$).  Differences between 
the results for the two mixtures are overall reasonably small, especially 
for ages above 1~Gyr.  For these old populations the largest color 
variations at a fixed age are $\sim$0.05~mag for $(J-K)$ and $(V-I)$, in 
the sense of the $\alpha$-enhanced SSPs (that have a higher $Z$ at fixed 
[Fe/H]) being redder. The $(U-B)$ colors at fixed [Fe/H] for the two 
metal distributions are overall more similar than the 
case of $(J-K)$ and $(V-I)$.

\begin{figure}
\plotone{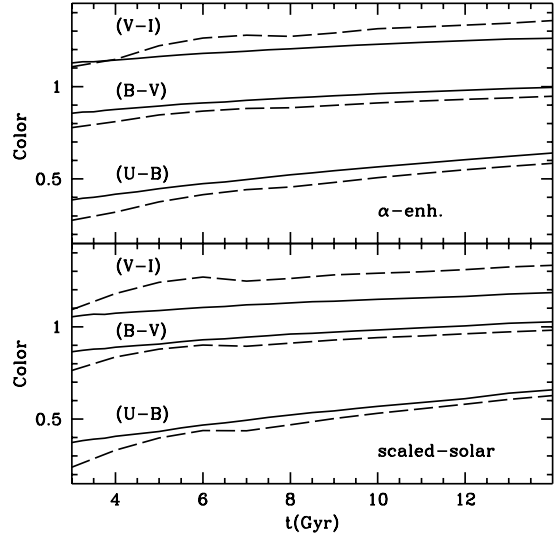}
\caption{Comparison of BaSTI colors (solid lines) with those from 
\citet{coelho07} (dashed lines).  The BaSTI colors are from our scaled 
solar [Fe/H]=+0.06 and $\alpha$-enhanced [Fe/H]=+0.05 models.  
The Coelho models are both for [Fe/H]=0.0.}
\label{colcomp}
\end{figure}

We can also compare our integrated colors to the predictions from the very 
recent models by \citet{coelho07}. They computed integrated spectra and 
colors for a range of SSPs with scaled solar and $\alpha$-enhanced metal 
mixtures using a method which is qualitatively similar to ours. Their 
choice for the $\alpha$-enhancement is essentially equal to ours, but the 
stellar isochrones and spectra employed in their modelling are different, 
and they also cover only a restricted metallicity range at present, with 
[Fe/H] between $-$0.5 and +0.2.
Their [Fe/H] gridpoint closest to ours is [Fe/H]=0.0 for both scaled-solar 
and $\alpha$-enhanced mixtures, and so we can compare these to our scaled 
solar [Fe/H]=+0.06 and $\alpha$-enhanced [Fe/H]=+0.05 models. 
Figure~\ref{colcomp} displays the $(U-B)$, $(B-V)$ and $(V-I)$ integrated 
colors for the common age range. Our colors are systematically redder
for both $\alpha$-enhanced and scaled-solar mixtures, except for the $(V-I)$ 
colors, where the reverse is true.  Typical differences are of the order 
of $\sim$0.1~mag.  In section~\ref{sec:line}, which compares the behaviour 
of line indices predicted by the 2 sets of models (ours and 
\citealt{coelho07}), we discuss some factors which may contribute to 
these differences.

\subsection{Integrated spectra: high resolution}
\label{sec:hrs}

Our high resolution integrated spectra are suitable for deriving line 
strengths (e.g. of Lick-style indices) and studying their behaviour as 
a function of age, metallicity and $\alpha$-enhancement (see 
section~\ref{sec:line}).  We also use them to make some preliminary tests 
of the effects of, for example, horizontal branch morphology, and 
statistical sampling in resolved clusters (see 
section~\ref{sec:comparison}).  They are constructed using the library of 
synthetic spectra from \citet{munari}.  The \citet{munari} spectra were 
computed using the SYNTHE code by \citet{kur93} and the input 
model atmospheres 
are the same as those employed to calculate the low-resolution spectral 
library described above \citep{castkur} -- hence they have similar 
metallicity, $T_{eff}$ and log$g$ coverage for both scaled solar and 
$\alpha$-enhanced versions\footnote{see 
http://archives.pd.astro.it/2500-10500/ for details of coverage}.
The \citet{munari} high resolution spectra take into account the effects of 
several molecules, including
${\rm C_2}$, CN, CO, CH, NH, SiH, SiO, MgH, OH, TiO and ${\rm H_2O}$. The
source of atomic and molecular data is the Kurucz atomic and molecular
line-list \citep[see, e.g.][]{kur92} with the exception of TiO and ${\rm
H_2O}$ data, for which the line lists of, respectively, \citet{schwenke} and
\citet{partridge} were adopted.
The wavelength range of the Munari spectra is 2500--10500{\AA} and the 
authors provide several 
different resolutions -- we used those with a uniform dispersion of 1{\AA}/pix
here.  

The referee suggested that we should investigate the comparison
between the \citet{munari} theoretical stellar spectra and their empirical 
counterparts, particularly for cool stars, for which Arcturus is the best
studied example.  In fact, our adopted high-resolution spectral library
has been extensively tested against observed spectra by \citet{martco}. 
In particular, they measured 35 spectral indices defined in the literature 
on three synthetic spectral databases (including the \citealt{munari} 
library adopted in
this work) and compared them with the corresponding values measured on
three empirical spectral libraries, namely Indo-US \citep{indous},
MILES \citep{miles} and ELODIE \citep{elodie}. The reader is referred to this
very informative paper for details about the comparisons and results. As a
general conclusion, all three sets of synthetic spectra tend to show the
largest discrepancies with the empirical counterparts at the coolest
temperatures. However, the quantitative and qualitative trends of these
differences depend on the specific empirical dataset used for comparison.

We now consider briefly the H$\beta$ and Fe5406 Lick-style indices, which we 
will be using extensively in the rest of this paper (see Figs. 29, 35 and 
Tables C1, C2 and C3 in Martins \& Coelho 2007). For H$\beta$ the best 
agreement, as given by the $adev$ parameter defined in \citet{martco}, is 
with the Indo-Us library for $T_{eff}>$7000~K (high temperatures) and
$T_{eff} \leq$4500~K (low temperatures), whereas the best agreement at
intermediate temperatures is with the MILES library (the
reader is warned that, judging from Tables C1, C2 and C3, some of the $adev$
values in their Figs. 29 and 35 have been mistyped). For a solar-like
dwarf star, the \citet{munari} spectra reproduce well the observed value,
whereas for an Arcturus-like giant the theoretical H$\beta$
index is lower than observations.
In general, for all three temperature ranges there is no clear systematic
offset between theory and observations, but a non-negligible spread in the
difference between them.

In the case of Fe5406 the best agreement is with the MILES database at
intermediate and high temperatures, with ELODIE being only marginally more
in agreement than MILES at low temperatures. The Indo-US library displays a
clear systematic difference at low temperatures (theoretical indices being
larger) and, to a lesser extent, also in the intermediate regime. These
offsets are much less evident in comparisons with the ELODIE database. For a
solar-like
dwarf star our adopted spectra tend to slightly overpredict the observed
value, whereas for an Arcturus-like giant the theoretical
index is in agreement with observations from the ELODIE library, slightly
overpredicted compared to the Indo-US database,
and massively overpredicted compared to MILES.

The reasons for the discrepancies with observed spectra are certainly at
least partly due to inadequacies in the theoretical spectra, but, as noted
also by \citet{martco}, errors in the atmospheric parameters of
the observed stars may also play a crucial part, especially in explaining
the different trends with observations coming from the different empirical 
libraries.

To this purpose, we made our own test on Arcturus, using an empirical
spectrum from the ELODIE spectral library.  We used their 0.2{\AA} resolution 
spectrum, which we degraded to 1{\AA} resolution to match the resolution of the
\citet{munari} library.  We compared this empirical spectrum with a theoretical
one which we created by interpolating amongst the Munari spectra, using
stellar parameters from \citet{peterson} -- namely  [Fe/H]=$-$0.5 ($\pm$0.1), 
$T_{eff}$=4300 ($\pm$30~K) and  log$g$=1.5 ($\pm$0.15).  We used an
$\alpha$-enhanced theoretical spectrum since \citet{peterson} found 
$\alpha$-elements enhanced by $\sim$0.3 dex, although it should be noted that the
theoretical spectra are fixed at [$\alpha$/Fe]=0.4.

Measuring H$\beta$ and Fe5406 line strengths directly on these spectra (see 
section~\ref{sec:line} for more details on this), we found that H$\beta$ is
under-predicted in the synthetic spectrum by 0.25{\AA} (measuring 0.61{\AA}, 
whereas the empirical spectrum gives 0.86{\AA}), whilst the Fe5406 line is 
slightly over-predicted (2.27{\AA} cf. 2.18{\AA}).  
We noted that the Mg$b$ line is also overpredicted by a similar fraction.
An underpredicted value for the H$\beta$ index is most likely
explained by the lack of non-LTE and/or chromospheric contributions and/or
inadequate treatment of convection \citep[see, e.g.][] {martco,korn05} in 
the theoretical spectra. However, the effect of uncertainties in the 
atmospheric parameters determined from empirical spectra may play 
also a crucial part in explaining these differences.  
As a test, we  modified $T_{eff}$ and log$g$ in the synthetic Arcturus
comparison spectrum to see the effect on derived indices.  We found that by
increasing $T_{eff}$ by 70K and decreasing log$g$ by 0.25 dex
($\sim$ 2$\sigma$ change in both parameters) the H$\beta$ line strength 
increased to 0.83{\AA} and Fe5406 decreased to 2.12{\AA}, both within 3\% of 
the empirical values.  The Mg$b$ line strength also decreased to come into
closer agreement with the empirical spectrum.

The full integrated SED derived from the high resolution spectra differs 
from that of the low resolution spectra because the \citet{munari} spectra 
do not include temperatures cooler than $T_{eff}=3500$K (the low 
temperature limit of the ATLAS9 model atmospheres) and hence the full SED 
differs from that of the low resolution spectra, particularly longward 
of $\sim$ 6000{\AA}.  This means that our high resolution integrated spectra 
are, in general, not suitable for deriving broad band 
colors and hence we use the low resolution versions for this. 
Figure~\ref{sample_spec} shows some representative examples of our 
high resolution integrated spectra.

\begin{figure}
\plotone{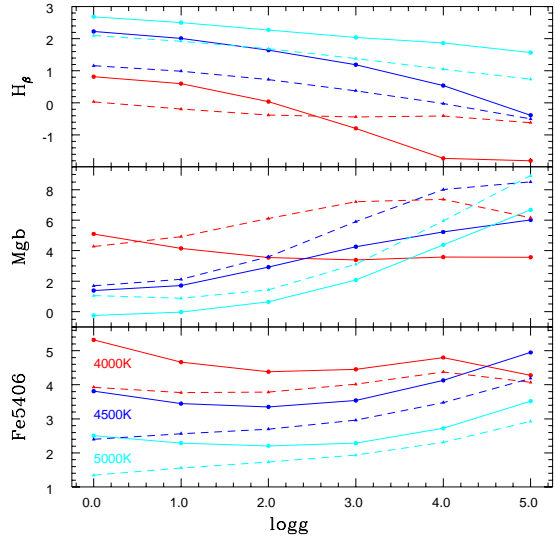}
\caption{Comparison of MARCS (solid lines) and Munari (dashed lines) models 
(scaled solar, [Fe/H]=0.0) for H$\beta$, Mg$b$ and Fe5406, at effective 
temperatures $T_{eff}$=4000K (red), 4500K (blue), 5000K (cyan).}
\label{marcs_comp}
\end{figure}

We decided not to attempt to include (or create) lower temperature 
high resolution spectra because we are concerned about possible 
inconsistencies between stellar atmosphere models from different sources.  
Figure~\ref{marcs_comp} shows a comparison between line strengths derived 
from the \citet{munari} spectra (based on ATLAS9 model atmospheres) and
corresponding spectra from the MARCS 
database\footnote{http://marcs.astro.uu.se/} -- we compare strengths
of the H$\beta$, Mg$b$ and Fe5406 lines since we will be using these 
extensively in our more detailed analyses later in this paper.  
It can be seen that for all 3 lines there are significant offsets, and 
that the behaviour becomes increasingly discrepant between the 2 models 
as the temperature decreases.  We note
that these comparisons were done using the MARCS models available as of May 
2008.  New low temperature MARCS models are currently being produced 
\citep{marcs} -- we will investigate these as they become available and 
may include them in our integrated spectra at a later date.
We note here that this low temperature regime also includes cool
AGB stars, for which no realistic models currently exist.

\begin{figure}
\plotone{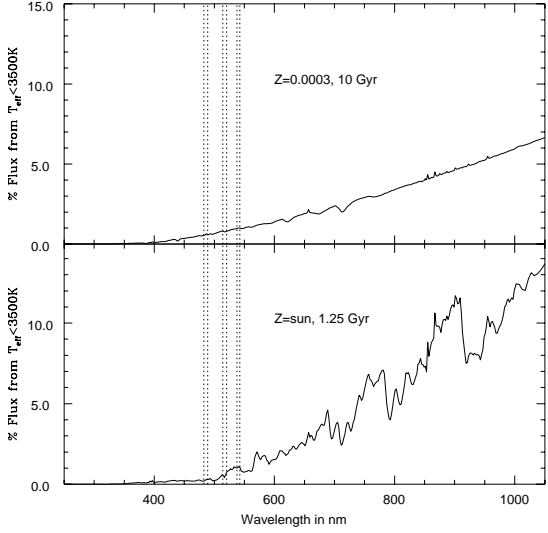}
\caption{Percentage of flux in stars with $T_{eff} <$ 3500K for 2 
representative SSPs.  Bandpasses for H$\beta$, Mg$b$ and Fe5406 are marked 
with the dotted lines.}
\label{fluxlost}
\end{figure}

The effect of not including stars with $T_{eff} < 3500$K in the high 
resolution integrated spectra is illustrated in Figure~\ref{fluxlost}.  
Here we use the low resolution spectra to separate out the flux contribution 
from these stars, for two representative SSPs, and plot the percentage of the 
total flux coming from $T_{eff} < 3500$K as a function of wavelength.  
It can clearly be seen that the 
flux contribution from these stars only becomes significant at wavelengths 
longer than the bandpasses of the main Lick-style indices, e.g. H$\beta$, 
Mg$b$ and various Fe lines (see \ref{sec:line}).

It is expected that high metallicity SSPs will be most affected by this
low temperature cut-off in the spectral library, since a larger portion of
their isochrones fall in the $T_{eff} < 3500$K regime than for low metallicity
SSPs.  As a test of the maximum likely effect on various diagnostic lines 
(see next section for definitions), we mimicked the effect of a 
$T_{eff} = 3500$K cut on a $Z$=0.04 (i.e. $\sim$ twice solar), 10~Gyr 
SSP by taking a 10~Gyr, $Z$=0.001 ($\sim \frac{1}{20}$ solar) isochrone 
and imposing a temperature cut at $T_{eff} = $4130K.  This temperature 
cut corresponds to approximately the same evolutionary point on the isochrone 
as a $T_{eff} = 3500$K cut for the higher metallicity SSP.  We then created
an integrated spectrum for this `truncated' $Z$=0.001 SSP and compared the 
derived line indices with those derived from the integrated spectrum for 
the full SSP at the same age and metallicity.  
For the `truncated' spectrum, the H$\beta$ index (the main age indicator -- 
see next section) was $\sim$ 0.1 higher than that of the full spectrum.  
This corresponds to an age difference of $\sim$ 1 Gyr, and goes in the sense 
that the truncated spectrum looks too young.  There are various metallicity
indicators (see next section) -- for the main [Fe/H] indicator we use here,
Fe5406, the line strength is slightly lower than for the full spectrum, 
which would result in an underestimate of the true [Fe/H] by $\sim$ 0.05 dex.
We repeated this test for a 3~Gyr model and found similar absolute offsets,
which would result in slightly smaller effects on the inferred age, at the 
level of $\sim$ 0.5~Gyr.

It should be noted that the issue of `missing flux' at low temperatures, 
and its associated effect on line indices, is not a problem specifically 
associated with the use of theoretical spectral libraries. It is also 
very relevant to any population synthesis techniques which rely on 
empirical spectra, since there are very few local stars at these low 
temperatures, particularly at low metallicity.


\subsubsection{Line indices: behaviour as a function of age and metallicity}
\label{sec:line}


All the line indices discussed in this paper are defined by the bandpasses 
tabulated in \citet[i.e. the 21 classical Lick/IDS indices]{trager98}. 
The quoted line strengths, and hence those displayed in plots, were 
obtained using the LECTOR programme by 
A. Vazdekis\footnote{See http://www.iac.es/galeria/vazdekis/index.html} and  
are those directly measured on the spectra, i.e. they are {\it not} 
transformed onto the Lick system.

Before discussing the behaviour of various line indices it is worth 
remembering the distinction between iron abundance, quantified here as [Fe/H],
and total metallicity, $Z$.  For scaled solar models, all metal abundances 
scale as for the sun, whilst in $\alpha$-enhanced models the $\alpha$ 
element abundances are increased relative to Fe.  Hence at fixed $Z$, 
the $\alpha$-enhanced models have a {\it lower} [Fe/H] than the 
corresponding scaled solar models.  As an example, the BaSTI scaled 
solar models with $Z$=0.0198 (the solar value) have [Fe/H]=+0.06, whilst 
the corresponding $\alpha$-enhanced models have [Fe/H]=$-$0.29.  
In the following discussion we will focus in detail on specific 
line indices which we will be using diagnostically to determine iron content 
[Fe/H], total metallicity $Z$, and age. 

\begin{figure}
\plotone{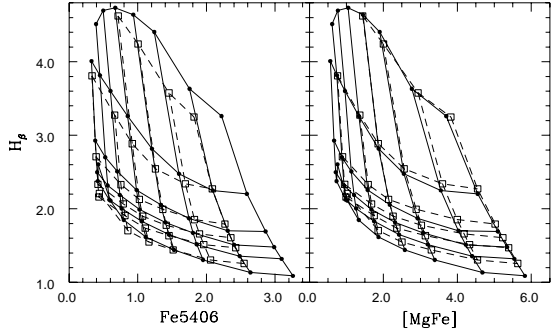}
\caption{Grids of H$\beta$ vs. Fe5406 (left panel) and [MgFe] (right panel) 
measured on our high resolution spectra as described in the text.  The solid 
line/circles are the scaled solar grid, the dashed line/open squares are the
$\alpha$-enhanced grid. Scaled solar models are for $Z$=0.0003, 0.0006, 0.001, 
0.002, 0.004, 0.008, 0.0198, 0.04 (corresponding to [Fe/H]=$-$1.79, $-$1.49, 
$-$1.27, $-$0.96, $-$0.66, $-$0.35, +0.06, +0.40) and $\alpha$-enhanced models 
are for $Z$=0.0006, 0.002, 0.004, 0.008, 0.0198, 0.04 (corresponding to 
[Fe/H]=$-$1.84, $-$1.31, $-$1.01, $-$0.70, $-$0.29, +0.05).  Metallicity
increases from left to right.  Ages (increasing from top to bottom) are
1.25, 3, 6, 8, 10 and 14 Gyr.}
\label{grid1}
\end{figure}

The left-hand panel of Figure~\ref{grid1} shows grids of H$\beta$ vs. Fe5406 
derived from our SSP integrated spectra for a range of metallicities and ages, 
as detailed in the figure caption.  The figure includes both the scaled 
solar and $\alpha$-enhanced grids and clearly demonstrates that the Fe5406 
line is predominantly a tracer of iron abundance, [Fe/H], since the
2 grids closely correspond along lines of very similar [Fe/H], and is 
generally insensitive to the degree of $\alpha$ enhancement.  This is not the 
case for other Fe line indices we have tested, which also include other 
dominant elements, as listed in \citet[][their Table 2]{trager98}.  
Although our line indices are not converted onto the Lick system, this 
sensitivity of the Fe5406 line to Fe only is similar to that noted in 
other studies which do utilize indices on the Lick system, e.g. 
\citet{leeworth} and \citet{korn05}.

The right-hand panel of Figure~\ref{grid1} shows similar grids for 
H$\beta$ vs. [MgFe], where [MgFe] is defined as 
$\sqrt{{\rm <Fe>} \times {\rm Mg}b}$ and
$<$Fe$>$=$\frac{1}{2}$(Fe5270+Fe5335).  It can clearly be seen that [MgFe] 
is sensitive to total metallicity, $Z$, since the scaled solar and 
$\alpha$-enhanced grids almost exactly correspond along lines of constant 
$Z$ across a wide metallicity range.  An important trend to note in 
this plot is the behaviour of the H$\beta$ line which, at fixed total 
metallicity $Z$ and age, is {\it stronger} in the $\alpha$-enhanced 
models than the scaled solar ones.  This behaviour of the H$\beta$ line 
strength is qualitatively similar to that found by \citet{tantchi07}, 
who used high resolution spectra to determine response functions which are 
used in conjunction with existing fitting functions to `correct' solar 
scaled indices to the appropriate $\alpha$-enhancement (effectively, an 
update of the method used by \citealt{tripbell}). 
However we note here that the recent work of \citet{coelho07}, discussed 
in Section~\ref{sec:cols}, seems to display the opposite trend in the 
behaviour of H$\beta$ with respect to $Z$ (see discussion below).

These 2 figures seen together (i.e. the 2 panels of Figure~\ref{grid1}) 
illustrate very well the difficulty in predicting ages from the H$\beta$ 
line without also having other information, in particular, the degree of 
$\alpha$ enhancement.  From our models it seems that the H$\beta$--Fe5406 
combination can be used to unambiguously determine age and [Fe/H], but 
yields no information about the degree of enhancement.  Conversely, 
using H$\beta$--[MgFe] alone could result in spurious ages unless the 
appropriate level of $\alpha$ enhancement is determined by other means.  
Hence the diagnostic use of a single index-index diagram is not recommended.

\begin{figure}
\plotone{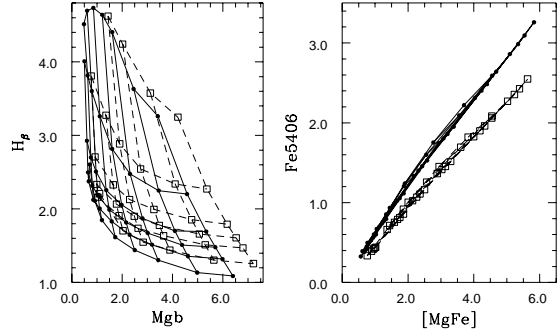}
\caption{H$\beta$ vs. Mg$b$ (left panel) and Fe5406 vs. [MgFe] (right panel). 
Models and symbols are as for Figure~\ref{grid1}.}
\label{grid2}
\end{figure}

In Figure~\ref{grid2} (left-hand panel) we show the corresponding grids
for H$\beta$ vs. Mg$b$ which demonstrate the sensitivity of the Mg$b$ index
to the degree of $\alpha$ enhancement (note that in our models this is fixed
at $[\alpha$/Fe]=0.4).  Qualitatively similar behaviour is also seen by 
\citet{schi07}.  However, this is a difficult grid to use diagnostically 
because the sensitivity to changing [Fe/H] and $Z$ are impossible to 
disentangle from this grid alone and hence, in our later analysis, we 
will utilise the Fe5406--[MgFe] plane, illustrated in the right-hand panel 
of Figure~\ref{grid2}.  In this plane, the lines of constant age are almost 
completely degenerate, however the scaled solar and $\alpha$-enhanced grids 
are clearly separated.  Using a combination of 3 grids -- H$\beta$ vs. Fe5406,
H$\beta$ vs. [MgFe], and Fe5406 vs. [MgFe] -- it should be possible to 
disentangle age, total metallicity $Z$, [Fe/H] and the degree of $\alpha$ 
enhancement, as we demonstrate in Section~\ref{sec:47tuc}.

\begin{figure}
\plotone{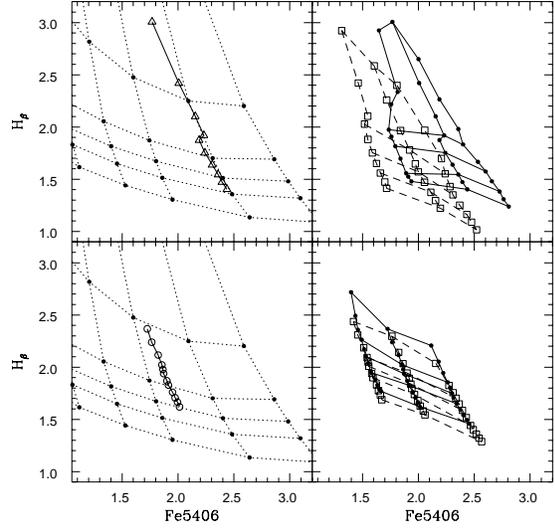}
\caption{Comparison of our H$\beta$ vs. Fe5406 grids with those of
\citet{coelho07} (upper panels) and \citet{tmb03} (lower panels).  
The left panels show the Coelho (upper) and Thomas (lower) 
[Fe/H]=0.0, scaled solar models 
for ages 3-12~Gyr superimposed on our scaled solar, H$\beta$ vs. Fe5406 grid.  
Upper right shows the Coelho grids for scaled solar (solid) 
and $\alpha$-enhanced (dashed lines) for  [Fe/H]= $-$0.5, 0.0 and +0.2. Lower
right shows TMB03 scaled solar grids for [Fe/H]=$-$0.33, 0.0, 0.35 (solid) 
and $\alpha$-enhanced ([$\alpha$/Fe]=0.3), [$Z$/H]=0.0, 0.35, 0.67, 
corresponding to [Fe/H]=$-$0.28, 0.06, 0.38  (dashed).}
\label{c07_comp}
\end{figure}

We compared the line strengths derived from our high resolution integrated
spectra to those of \citet{coelho07}, a recent study which uses a method most
similar to ours to produce integrated spectra for a range of SSPs (see 
Section~\ref{sec:cols} for a comparison with their colors).  The upper left 
panel of Figure~\ref{c07_comp} shows our scaled solar H$\beta$ vs. Fe5406 grid 
with the Coelho data superimposed for [Fe/H]=0.0 (ages 3-12~Gyr).  There 
is reasonably good agreement between their models and ours
in the general locus of points along the [Fe/H]=0.0 line, however there
are obvious offsets in H$\beta$ which would cause a discrepancy in predicted 
ages.  In general, their scaled solar grid lies at higher H$\beta$ values 
than ours and so for a fixed (observed) data point we would predict younger 
ages.  The upper right hand panel of Figure~\ref{c07_comp} shows the full grid 
of H$\beta$ vs. Fe5406 from the Coelho models (note, only 3 metallicities 
are currently available) -- this figure seems to show the opposite trend
in the behaviour of $H\beta$ with respect to $\alpha$-enhancement compared
to our models, as mentioned above.  

We have no obvious explanation for this difference in behaviour but it 
should be noted that, although \citet{coelho07} use a method which is 
qualitatively similar to ours to produce integrated spectra, they are 
using different underlying stellar models and a different spectral library. 
In particular, we note here that the isochrones used in the Coelho models
differ signficantly from the BaSTI ones when similar ages and metallicities
are compared.  In general, their RGBs are at cooler temperatures than the 
BaSTI ones, by up to $\sim$ 200K.  Also the temperature at the turn off (TO) 
displays significant differences in behaviour, e.g. at old ages, the BaSTI
isochrones have hotter TOs, whilst at intermediate and young ages this 
trend is reversed, and the BaSTI TOs are cooler.
\citet{coelho07} also include some low temperature spectra in their models,
generated using the MARCS model atmospheres, in order to cover all 
evolutionary stages -- this enables them to compute broadband colors from 
their high 
resolution spectra, whilst we use our low resolution spectra for this.  
They also compute models for different values of total metallicity $Z$ to 
ours, meaning that a comparison of grids (i.e. with a range of ages and 
metallicities) is very difficult -- only their scaled solar, [Fe/H]=0.0 
models are directly comparable with ours.

For the interested reader, we also include a similar comparison with the
fitting-function based models of \citet{tmb03} -- these are displayed in
the lower panels of Figure~\ref{c07_comp}.  By definition, fitting-function
based models are on the Lick system and so we have modified the tabulated
H$\beta$ and Fe5406 values by 0.13 and 0.2 respectively, according to the 
prescription of \citet[][their Table~6]{bc03}, to make them directly
comparable with indices derived from flux calibrated spectra, as used in 
our work.  

We note here that all of our models presented so far in this paper have used 
the BaSTI non-overshooting models with a 
value for the Reimers mass loss parameter of $\eta$=0.2.  The effect of 
varying $\eta$ will be discussed in the next section.

\section{Comparisons with resolved stellar clusters}
\label{sec:comparison}

Any stellar population synthesis model should be tested 
against resolved populations with independent estimates of chemical 
composition and age.  Star clusters belonging to the Galaxy are the 
obvious choice, because their chemical composition can be determined 
from spectroscopy, and theoretical isochrones matched to the observed 
color-magnitude diagram (CMD) provide an estimate of the age.  
Ideally, the analysis of integrated 
colors and spectra by means of population synthesis methods should
provide a chemical composition and age consistent with the inferences from
spectroscopy and the CMD
\citep[see, e.g.][and references therein]{gibson99,vaz01,schi07}.

To this purpose one can make use of the \citet{schi05} library of
integrated spectra of 40 Galactic globular clusters, plus the integrated
spectrum of the Galactic open cluster M67 \citep{schi04}.
Each of the globular cluster integrated spectra covers the range
$\sim 3350 - 6430${\AA}, with $\sim$ 3.1{\AA} (FWHM) resolution, and one 
can potentially attempt fits to the whole sample of spectra, comparing 
the results derived from diagnostic lines to those derived from isochrone
fitting to the observed CMD.  There are however some
serious issues that one has to take into
account when interpreting the results of these comparisons.

The first problem is the [Fe/H] estimates of the clusters. There are large
discrepancies between the widely employed \citet{zinn} and \citet{cg97} 
scales.  For example, well studied objects like M3 and M5 show differences 
of $\sim$0.3~dex between the two [Fe/H] estimates.
A third [Fe/H] scale by \citet{krafti} based on FeII lines shows
very large differences compared to the previous two sets of estimates in 
the low metallicity regime.
As an example, a well studied metal poor cluster like M68 displays a $\sim$
0.4~dex range of [Fe/H] values when these three different [Fe/H] scales are
employed.  Taking a mean value of the various determinations for each cluster
does not make much sense, given that the differences among the authors are
mainly systematic. In addition, [$\alpha$/Fe] spectroscopic estimates 
do not exist for many clusters.

The second issue is the presence of the well known CN and ONa 
anticorrelations in the metal mixture of stars within a single cluster 
\citep[see, e.g., the review by][and references therein]{gratton04}. 
There is now convincing evidence that C, N, O, Na -- and sometimes
also Mg and Al -- display a pattern of abundance variations superimposed 
onto a normal $\alpha$-enhanced heavy element distribution.
Negative variations of C and O are accompanied  by increased N and Na
abundances. There is also now general agreement that these abundance 
patterns are of primordial origin.
This means that the integrated spectrum of a cluster is composed of 
individual spectra with a range of values for the very important CNONa 
elements, whereas the Fe abundance is essentially constant
for all stars. It is therefore very difficult to interpret the strength of
all line indices affected by these 4 elements, given that in a cluster
the exact proportions of stars with a certain
set of CNONa abundances, and their location along the CMD, are known
only for small samples of objects.

The third issue is related to the morphology of the horizontal branch (HB). 
As is well known \citep[see, e.g.][]{schi04b}, the presence of blue HB stars
affects the Balmer lines, which can produce spuriously young ages for clusters 
with an extended blue HB if this is not included in the theoretical modelling.
A detailed synthetic modelling of the HB color extension for each of the 
observed clusters is therefore in principle necessary.  Whilst this is 
possible in principle, it is difficult to implement in practice and would 
be computationally very time consuming to model all the different 
possibilities.  This approach also assumes that one knows, a priori, the 
appropriate morphology required -- this of course is not possible for 
unresolved stellar clusters.  The color of the HB is determined by the 
mass loss history of RGB stars when the initial chemical composition is fixed.
In the theoretical isochrones, this is modelled using the Reimers mass loss 
law, and the mass loss parameter $\eta$ is a free parameter in the models,
which is set at some fixed value for each isochrone.  This means that for 
any particular isochrone, all the stars on the HB have essentially the 
same mass and hence start their evolution from the Zero Age HB all with 
the same color. For any fixed $\eta$ value, higher metallicity SSPs have 
a red clump of stars along the HB. The distribution of objects along the HB 
becomes progressively bluer when the metallicity decreases, provided 
that $\eta$ is unchanged.  
In Galactic globular clusters however, the mean color and color 
extension of the HB is often not correlated to the cluster metallicity, and
additional chemical or environmental parameters that directly or indirectly
affect the mass distribution along the HB, must be at play 
\citep[see e.g.][and references therein]{caloi,recio}.
If $\eta$ is fixed at a higher value in the models, the blue 
extension of the HB becomes more extreme at low metallicity, particularly for 
old ages (see discussion below).

A final issue is the existence of blue stragglers in the CMDs of Galactic
globular clusters, i.e. a plume of stars brighter and generally bluer 
than the TO, thought to be created through mass transfer in a binary 
system or through direct collision of 2 stars.
Integrated spectra that include only the contribution
of single star evolution miss this component, causing potential biases in
their age estimates.

Keeping in mind all these limitations, we first compare the whole globular
cluster (GC) spectral library by \citet{schi05} to our high resolution spectra 
calculated with both $\eta=0.2$ and $\eta=0.4$.  We have measured indices 
on the GC spectra in exactly the same way as for our model spectra.  We note 
here that the \citet{schi05} data set includes multiple observations for 
several clusters -- in the following plots all the data points are shown.  
To minimize the effect of the CNONa anticorrelations on the comparison,
we consider first the Fe5406$-$[MgFe] plane to estimate [Fe/H] 
(which is constant among all stars in a cluster) and the degree of 
$\alpha$-enhancement from our theoretical spectra. As seen in 
section~\ref{sec:line}, [MgFe] is sensitive to the global metallicity of the 
stellar population, but not to the exact distribution of the metals, 
and therefore the Fe5406$-$[MgFe] diagram -- insensitive to age for 
old stellar populations -- depends on the relationship between
the global metallicity $Z$ and the iron content in the population. Given
that at fixed [Fe/H] an $\alpha$-enhanced mixture has a larger $Z$ 
compared to a scaled-solar one, the position of a cluster in this diagram 
is related to the degree of $\alpha$-enhancement in the spectra of its stars. 
A possible complication is that the CNONa anticorrelation modifies the baseline
$\alpha$-enhanced metal mixture in a fraction of stars within the cluster.
However, if the sum of C+N+O (which makes up a substantial fraction of the 
total metals) is unaffected by the abundance anomalies (as seems to be the 
case within current observational errors, see, e.g. \citealt{carretta05}) 
the relation between [Fe/H] and $Z$ is unchanged, and the Fe5406$-$[MgFe] 
diagram still provides an estimate of the degree of $\alpha$-enhancement 
in the baseline chemical composition of the cluster as a whole.

\begin{figure}
\plotone{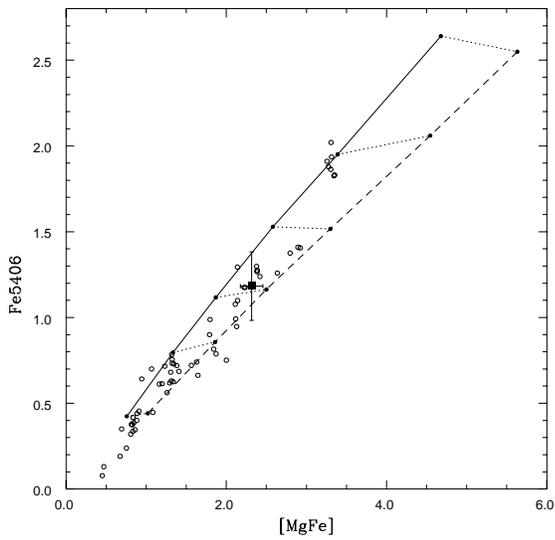}
\caption{Fe5406 vs. [MgFe] from our scaled solar and $\alpha$-enhanced 
14~Gyr spectra
(solid and dashed lines, respectively) joined at approximately equal [Fe/H] 
(dotted lines) for [Fe/H] $\sim$ +0.06, $-$0.3, $-$0.7, $-$1.0, $-$1.3, $-$1.8 
(decreasing from top to bottom).  Open circles are the Galactic globular 
cluster data from \citet{schi05} -- 47~Tuc is represented by the solid square, 
with an estimated typical error bar.}
\label{f5406_mgfe_ggc}
\end{figure}

Figure~\ref{f5406_mgfe_ggc} shows all the Schiavon data points in the 
Fe5406--[MgFe] plane, with lines of 14~Gyr (constant) age derived
from our scaled solar and $\alpha$-enhanced spectra, in the metallicity range
$-1.8<{\rm [Fe/H]}<+0.06$.  In general, the data points fall exactly between 
the scaled solar and $\alpha$-enhanced lines indicating that most of them 
are probably $\alpha$-enhanced to some extent, as expected.  The group of 
points at the top of the grid represents 2 clusters only -- NGC6528 (for 
which there are 6 observations) and NGC 6553, which are the 2 most metal 
rich clusters in the data set at [Fe/H]$\sim -0.2$.  The 4 points at the 
bottom left of the grid represent 2 observations each of NGC~2298 and 
NGC~7078, the 2 most metal poor clusters at [Fe/H]$\sim -2.0$.  Hence we 
maintain that this grid is a good first indicator of [Fe/H], and also 
provides an estimate of the degree of $\alpha$ enhancement.

\begin{figure}
\plotone{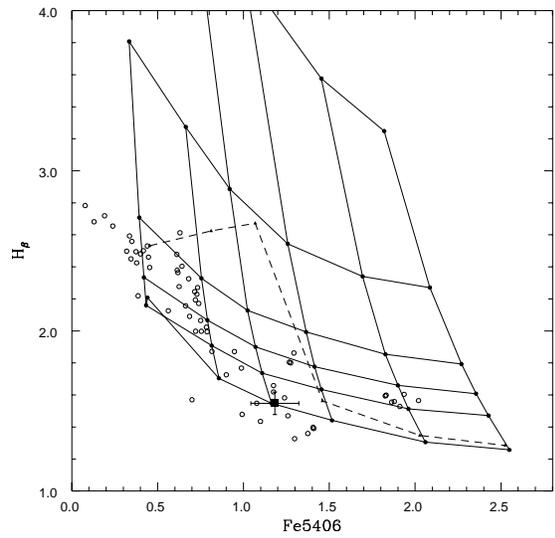}
\caption{H$\beta$ vs. Fe5406 grid for our $\alpha$-enhanced models 
([Fe/H]=$-$1.84, $-$1.31, $-$1.01, $-$0.70, $-$0.29, +0.05)
with GGC data from \citet{schi05} overplotted -- symbols as for 
Figure~\ref{f5406_mgfe_ggc}.  The grid shown is calculated using mass loss 
parameter $\eta$=0.2.  The dashed line is for the 14~Gyr models only and
was calculated using $\eta$=0.4, this illustrates the effect of a very 
extended blue horizontal branch on the predicted H$\beta$.}
\label{hb_f5406_ggc}
\end{figure}

Once [Fe/H] and the degree of $\alpha$-enhancement are established we can 
then use the H$\beta$--Fe5406 plane to estimate the cluster ages -- 
Figure~\ref{hb_f5406_ggc} shows the same GC data points on our 
$\alpha$-enhanced grid, calculated using $\eta$=0.2.  Whilst the 
intermediate metallicity clusters generally span the 10-14~Gyr age range 
(within the errors), it can be seen that the more metal poor clusters all 
seem to scatter to higher ages.  This is most likely to be due to 
the blue extent of their HBs, as discussed above.  To illustrate the potential
effect of having extended blue HBs due to increased mass loss on the RGB, 
we have overplotted the 14~Gyr line calculated using $\eta$=0.4 -- this 
clearly demonstrates the large increase in the strength of the H$\beta$ 
line which can lead to spuriously young ages.  We note that the trend of 
this $\eta$=0.4 line,
i.e. the rise to a peak in H$\beta$ at some intermediate metallicity and then a
decrease towards lower metallicities, is not unexpected.  For the most metal 
poor old clusters, the horizontal branch can develop a very long blue tail 
which extends down to magnitudes almost as faint as the turn off in optical 
bands.  Thus although the individual stars 
still exhibit strong H$\beta$ due to their very hot temperatures, their 
contribution to the overall continuum flux decreases at the wavelength 
of H$\beta$ (which is in the middle of the $B$-band) at the lowest 
metallicities.

\subsection{Case study: 47 Tuc}
\label{sec:47tuc}


We have performed a more detailed comparison for the case of 47~Tuc, which is
one of the clusters in the \citet{schi05} database.
This cluster has a reasonably well established [Fe/H]. \citet{cg97}
find [Fe/H]=$-0.70\pm 0.07$, \citet{zinn} give [Fe/H]=$-0.71\pm0.08$, while
\citet{krafti} obtain between $-$0.78 and $-$0.88 depending on the
assumed $T_{eff}$ scale, based on the equivalent widths of FeII 
lines measured from high-resolution spectra of giants.
More recent high resolution spectroscopy by \citet{carretta04} gives an
estimate consistent with that of \citet{cg97}, while
\citet{koch} obtain [Fe/H]=$-0.76\pm 0.01 \pm 0.04$.
Measurements of $\alpha$-element abundances by \citet{carretta04} and
\citet{koch} give $\alpha$ enhancement, [$\alpha$/Fe], of the order of 
$\sim0.3-0.4$~dex on average.

\begin{figure}
\plotone{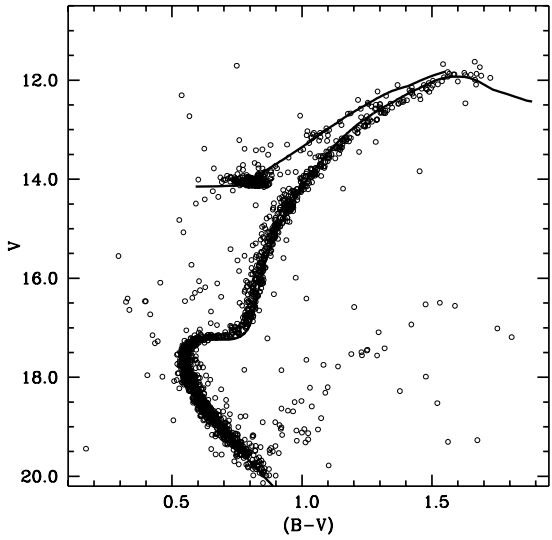}
\caption{Isochrone fit to 47~Tuc \citep[data from][]{stetson}.  The best 
fit to a 12~Gyr, [Fe/H]=$-$0.7, $\alpha$-enhanced isochrone yields a 
distance modulus of $(m-M)_{0}$=13.28, for an assumed reddening of 
$E$(B-V)=0.02, in agreement with the empirical main-sequence fitting result
of \citet{perci02}.}
\label{47tuc_fit}
\end{figure}

Figure~\ref{47tuc_fit} displays a fit to the $BV$ data by \citet{stetson}
with the BaSTI $\alpha$-enhanced, $\eta$=0.2 
isochrone and ZAHB, with [Fe/H]=$-$0.7 and an age of 12~Gyr.
A more formal determination of the age and associated errors has been
obtained using the $\Delta V$ method described in, e.g., \citet{sw97,sw98}. 
This method employs the difference in $V$ magnitude between the Zero Age 
HB (ZAHB) and the turn off (TO), designated $\Delta V$, as an age indicator. 
This quantity turns out 
to be weakly sensitive to the cluster metallicity, so that errors on the 
age estimate are dominated by errors on the determination of the ZAHB 
and TO magnitudes (see the discussion in \citealt{sw97}).
We obtain for 47~Tuc an age $t=12\pm$1~Gyr where in the error budget we have
included the small contribution of a $\pm$0.2~dex uncertainty around 
the reference [Fe/H]=$-$0.7.

A comparison of the 47~Tuc line indices with those derived from the
theoretical spectra (computed with $\eta$=0.2) in the Fe5406$-$[MgFe] 
and H$\beta-$Fe5406 diagrams is included in 
Figures~\ref{f5406_mgfe_ggc} and \ref{hb_f5406_ggc}.  We have estimated 
representative error bars for 47~Tuc, which will be approximately the same 
for all the GC data points.  These include contributions from the 
measurement of the line indices on the observational data as given by the 
LECTOR program, an estimate of the intrinsic observational error (there is 
only one observation of 47~Tuc), and a small contribution due to the 
difference in resolution between the data and the models.
Within the error bars of the observed indices, our BaSTI integrated spectra
yield [Fe/H], [$\alpha$/Fe] and age estimates in agreement with the 
results from high resolution spectroscopy of individual stars and from 
the CMD age.

We have then checked the extent of possible biases introduced by the
theoretical representation of the cluster HB, which differs from
the observed one. To this purpose we have considered
the Fe5406, [MgFe] and H$\beta$ values obtained from a 14 Gyr old, 
$\eta$=0.2, [Fe/H]=$-$0.7 theoretical spectrum. We have
then computed an integrated spectrum for the same age and [Fe/H], 
but considering a synthetic HB (obtained as described in \citealt{ms07})
that precisely matches the color extension observed in 47~Tuc. The resulting
Fe5406, [MgFe] and H$\beta$ values are practically unchanged compared 
to the $\eta$=0.2 case -- the increase in H$\beta$ is $<$ 0.04, corresponding 
to an age difference of $<$ 0.5~Gyr, and well within the observational 
error, whilst Fe5406 and [MgFe] are negligibly changed.

As a second check we have estimated the effect of the blue stragglers
present in the cluster CMD.  According to the estimates by
\citet{piotto02}, the ratio of blue stragglers to the number of HB
stars is about 0.15,
one of the lowest fractions within the Galactic globular cluster system.
To determine an upper limit of their effect on the indices used in our
analysis, we have added to the 14~Gyr old [Fe/H]=$-$0.7, $\eta$=0.2 
integrated spectrum, the contribution of a clump of Zero Age MS stars 
1.5 mag brighter than the isochrone TO, which is the approximate position 
of the bluest and brightest blue stragglers in the $BV$ CMD by \citet{sills}. 
Their number is assumed to be 15\% of the number of HB stars predicted by
the models. All line indices measured on the resulting integrated spectrum
are practically unchanged compared to the reference values, with differences 
of no more than 0.01 in any of the diagnostic indices used here.

\subsection{Case study: M67}
\label{sec:m67}


As a second detailed test, we have considered the open cluster M67. 
\citet{schi04} have created a representative integrated spectrum for M67, 
built up from observations of spectra of individual stars in the cluster
(about 90 objects) that sample as well as possible the CMD from the lower
MS to the tip of the RGB and the He-burning (red clump) phase.  Some local
field stars are also included to better sample some regions of the cluster CMD
not well represented by the M67 data alone.  \citet{schi04} assigned a mass 
to each of the sample stars by using inputs from 
theoretical isochrones, and then the individual spectra were
coadded with weights given by an adopted IMF (they employ a
Salpeter IMF). The contribution of objects with masses below 
$\sim$0.85 $M_{\odot}$ (below the faintest objects whose spectrum is 
included in the \citealt{schi04} analysis) has been included by 
\citet{schi07} using inputs from theoretical models.  The corrections 
that need to be applied to line indices measured on their published M67 
spectrum to account for this (their Table 11) have been included in 
our analysis.  The contribution of blue stragglers is excluded from the 
spectrum we employ in the comparisons that follow.

\begin{figure}
\plotone{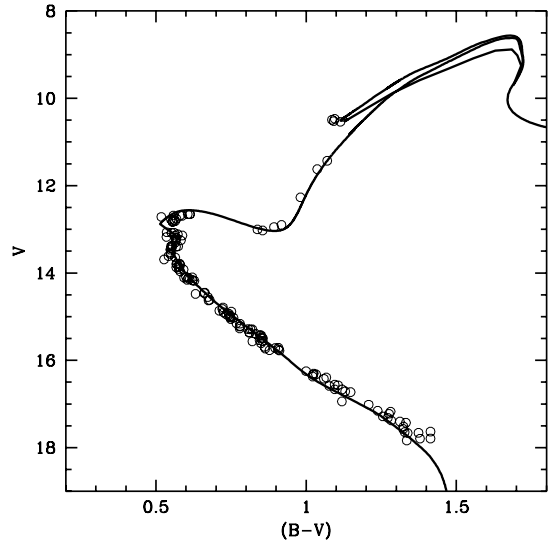}
\caption{Isochrone fit to M67 \citep[data from][]{sandquist}.  The best fit
to a scaled solar, non-overshooting, [Fe/H]=0.06 isochrone, yields an age of
4.0~Gyr and a distance modulus of $(m-M)_{0}$=9.67, for an assumed reddening 
of $E$(B-V)=0.02, in agreement with the empirical main-sequence fitting result 
of \citet{perci03}. }
\label{M67_fit}
\end{figure}

For this cluster the spectroscopic investigations by
\citet{taut}, \citet{yong}, \citet{randich} and the analysis by 
\citet{taylor}, all give [Fe/H]$\sim 0.0$.
The best age estimate obtained by fitting our [Fe/H]=+0.06 BaSTI 
non-overshooting isochrones to \citet{sandquist} photometry is $t$=4.0~Gyr. 
The inclusion of overshooting has only a marginal effect at this age, 
increasing $t$ to 4.2~Gyr.
The quality of the fit is very similar, with a small preference for the 
non-overshooting case, which we display in Figure~\ref{M67_fit}.

\begin{figure}
\plotone{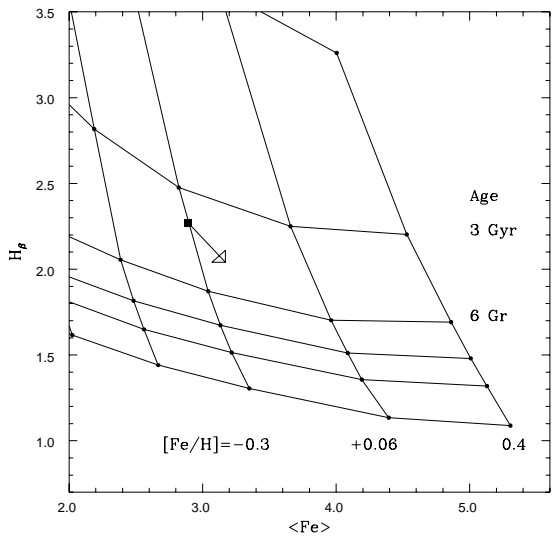}
\caption{Scaled solar grid for H$\beta$ vs. $<$Fe$>$ with the M67 data point, 
calculated as described in the text.  The arrow shows the approximate 
direction and distance the point would move if the effects of statistical 
fluctuations were taken into account.}
\label{hb_fe_m67}
\end{figure}

A comparison of the M67 line indices with those derived from our scaled solar 
non-overshooting grid (computed with $\eta$=0.2, but at the age of M67 the 
choice of $\eta$ is irrelevant) in the H$\beta-<$Fe$>$ diagram is displayed in 
Figures~\ref{hb_fe_m67}.  We used $<$Fe$>$ instead of the Fe5406 line here 
because the wavelength coverage of the M67 spectrum is slightly less than 
for the GC sample, cutting off at 5396{\AA}.  We have not assigned error bars 
to the data point on this plot as it not trivial to quantify them -- they 
must however be at least as large as the error bars estimated for the GC data, 
and some other potential contributory factors are discussed below.  
Whilst the age inferred from this diagram is consistent with the CMD age,
the predicted [Fe/H] is lower than the spectroscopic value by about 0.2~dex.  
There are some important factors which we think may contribute to this 
discrepancy which we address in the following discussion.

One important point to consider is that M67 spectrum does not come from
observations of the cluster integrated light (as is the case for the GC data 
set), rather it has been built in the same way as theoretical integrated 
spectra are built, starting from an isochrone.  Each point along the 
isochrone has been assigned an appropriate spectrum, and the
contributions of the individual spectra weighted by the chosen IMF. 
The BaSTI isochrones include more than 2100 points to sample masses above 
0.85$M_{\odot}$ -- a very fine coverage -- whereas the M67
spectrum has been computed using a much smaller number of individual
representative stars. To get a handle on the effect introduced by the 
sampling used for the observed spectrum, we have computed two integrated 
spectra for the same age of 4.0~Gyr ([Fe/H]=0.06, scaled solar 
non-overshooting models). 
The first one has been obtained as usual, whereas for the second spectrum 
the resolution of the isochrone for masses above 0.85$M_{\odot}$
has been degraded to include only 90 points. The spacing of these points has
been chosen to reproduce as well as possible the $V$ magnitude distribution 
of the stars used by \citet{schi04} (we have assumed a distance modulus 
$(m-M)_0$=9.67, as obtained from the isochrone fitting).
We find a non-negligible difference between line indices computed with these
two different isochrone samplings.
The value of the difference between the fine sampling and the coarse
sampling spectrum has then to be added to the measured values for M67, 
in order to have a more consistent comparison with the theoretical grid.
The direction in which the data point moves in the the H$\beta-<$Fe$>$ plane 
after applying this correction is indicated on Figure~\ref{hb_fe_m67} by the 
arrow.  The H$\beta$ value decreases, implying a slightly older age, 
although this may not be significant within the observational errors.  
Interestingly, all the Fe indicators move to significantly higher values, 
increasing the inferred [Fe/H] by as much as 0.2 dex and putting the 
predicted [Fe/H] more in line with the spectroscopic value.

This test highlights some subtle issues affecting the creation of SSPs which 
can have non-trivial effects on their derived integrated properties and 
spectra.  We remind the reader that the method of populating an isochrone 
to create an SSP used here (and in other population synthesis studies) is 
an analytical one, in which the isochrone is effectively treated as a 
continuous function.  This method is strictly correct only in the limit 
of an infinite number of objects in the stellar system being modelled, 
but is the appropriate method to use when modelling an unresolved population.
In practice, an isochrone consists of a discrete number of evolutionary points 
(EPs), all of which are assigned a spectrum when the analytical method is 
used.  However, the number of defined EPs, resulting in coarse or fine 
sampling of the isochrone, can make a difference to the resultant 
integrated properties, as demonstrated above.

A further refinement in the modelling of the integrated properties of observed
stellar clusters can be made by populating the isochrone with a discrete 
number of stars appropriate to the type of cluster (using an IMF, as before).
The Web tool provided on the BaSTI database which creates synthetic CMDs for 
resolved populations (SYNTHETIC MAN) populates isochrones in this way, 
with the number of stars in the simulation being determined by the user.  
When the number of stars in a particular simulation is not large enough 
to populate every EP on the isochrone (which is already true for typical 
$10^5-10^6~M_{\odot}$ Galactic globular clusters) statistical fluctuations 
of star counts will arise. 
This means that an ensemble of SSPs all with the same
total mass, age and initial chemical composition will display a range of
integrated magnitudes and colors, that increases (for a fixed value of 
${\rm M_t}$) when moving towards longer wavelengths, which are dominated 
by the shorter-lived (hence more sparsely populated) RGB and AGB 
phases.  Preliminary tests by us have shown that there are also non-negligible
effects on integrated spectra produced in this way and that a significant 
scatter is seen in derived line indices when the number of stars in the 
simulations corresponds to a total mass less than 
$\sim 2 \times 10^{5} M_{\odot}$ 
(i.e. roughly the size of a typical GC and larger than most open clusters).  
In particular, one significant factor seems to be the exact location of 
the brightest star populating the RGB, which can easily vary by several 
tenths of a magnitude when modelling open clusters like M67. 

All these issues, and particularly the effects of statistical fluctuations on 
integrated spectra and derived line indices, will be explored further in a 
subsequent paper.

\section{Summary}
\label{sec:sum}

In this 4th paper in the BaSTI series we have presented a new set of low 
and high resolution synthetic spectra based on the BaSTI stellar models, 
covering a large range of SSPs, for both scaled solar and $\alpha$ 
enhanced metal mixtures.  This enables a completely consistent study of 
the photometric and spectroscopic properties of both resolved and
unresolved stellar populations, and allows us to make detailed tests on
specific factors which can affect their integrated properties.

\begin{itemize}
\item
We have produced low resolution spectra suitable for deriving broadband 
magnitudes and colors in any photometric system and ensured that the 
derived colors are fully consistent with those predicted 
from the BaSTI isochrones.  

\item
We have produced high resolution spectra suitable for studying the behaviour 
of line indices and tested them against a large sample of Galactic globular 
clusters, making particular note of the effect of horizontal branch morphology.

\item
We have tested the global consistency of the BaSTI models by making detailed 
comparisons between ages and metallicities derived from isochrone fitting and 
from line index strengths for the Galactic globular cluster 47~Tuc and the 
open cluster M67.

\item
We find non-negligible effects on derived line indices caused by statistical 
fluctuations, which are a result of the specific method used to populate an 
isochrone and assign appropriate spectra to individual stars. 
\end{itemize}

\acknowledgments

We warmly thank the referee, Guy Worthey, for a constructive report and some
helpful comments and suggestions.
SC acknowledges the financial support of INAF through the PRIN 2007 grant 
n.CRA 1.06.10.04~~\lq{The local route to galaxy
formation: tracing the relics of the hierarchical merging process in the
Milky Way and in other nearby galaxies}\rq.
SMP acknowledges financial support from STFC through a Postdoctoral Research
Fellowship.  SMP would also like to thank PAJ for his endless patience and
encouragement during the preparation of this work.

\appendix
\section{Appendix}

Upon publication of this paper, the official BaSTI website
(\url{http://193.204.1.62/index.html}) will be expanded to include
integrated spectra (low and high resolution), magnitudes and mass-to-light
(ML) ratios for SSPs (normalized to 1$M_{\odot}$ of stars formed during 
the burst of star formation) covering the full range of ages,
chemical compositions, choices of mass loss and mixing included in the
database.  A full summary of the parameter space covered by BaSTI can be 
found in \citet{basti3}.
Integrated magnitudes and ML ratios will be provided for $UBVRIJHKL$ filters. 
Additional photometric systems can be easily accounted for by convolving 
the BaSTI low resolution integrated spectra
with the appropriate filter response curves, and using an appropriate
normalizing spectrum (a Vega synthetic spectrum is also included in BaSTI). 
ML-ratio files also tabulate
the amount of mass locked into the various types of remnants for each
SSP, to help determine ML-ratios
in any arbitrary photometric system.
We remind the reader that the ML-ratio of a stellar population is customarily
defined as the ratio between the integrated luminosity in a given passband
(in solar units) divided by the integrated stellar mass (in solar units)
contained in stars,
including all remnants like white dwarfs (WDs), neutron stars (NSs) and
black holes
(BHs). To compute the solar luminosity in the $UBVRIJHKL$ we have adopted
the following
set of solar absolute magnitudes:
$M_U=5.66, M_B=5.49, M_V=4.82, M_R=4.45, M_I=4.10, M_J=3.68, M_H=3.31,
M_K=3.28, M_L=3.26$.
The mass contained in WD, NS and BH remnants has been computed by assuming
an initial-final-mass (IFMR)
relation
in addition to the IMF. For WDs with a carbon-oxygen core we employed the
IFMR obtained from our AGB modelling.
The upper initial mass limit for the production of carbon-oxygen WDs (the
so-called $M_{up}$)
is also obtained from our stellar models.
For initial masses between $M_{up}$ and 10~$M_{\odot}$ we considered typical
oxygen-neon core WDs
with masses of 1.3$M_{\odot}$. Initial masses between 10 and 25~$M_{\odot}$
are assumed to give rise
to 1.4$M_{\odot}$ NSs, whereas for progenitors above 25$M_{\odot}$ we
consider a final state
as black holes with mass equal to $1/3$ of the initial mass.

\end{document}